\documentclass[reprint,
superscriptaddress,
bibnotes,
 amsmath,amssymb,
 aip,
apl,
showkeys,
floatfix,
]{revtex4-1}

\usepackage{placeins}
\usepackage{graphicx}
\usepackage{dcolumn}
\usepackage[T1]{fontenc}
\usepackage[utf8]{inputenc}
\usepackage{graphicx}
\usepackage{float}
\usepackage{natbib}
\usepackage{hyphenat}
\usepackage{tabularx}
\usepackage{amsmath}
\usepackage{adjustbox}
\usepackage{longtable}
\usepackage{epstopdf}
\DeclareUnicodeCharacter{0301}{\'{e}}
\usepackage{bm}

\begin{document}

\preprint{APS/123-QED}

\title{Enhancement of in-plane anisotropy in  MoS$_{2}$/CoFeB bilayers}

\author{Vijayabaskaran Thiruvengadam}
\affiliation{Laboratory for Nanomagnetism and Magnetic Materials (LNMM), School of Physical Sciences, National Institute of Science Education and Research (NISER), HBNI,, P.O.- Bhimpur, Padanpur, Via-Jatni, 752050, India}

\author{Abhisek Mishra}
\affiliation{Laboratory for Nanomagnetism and Magnetic Materials (LNMM), School of Physical Sciences, National Institute of Science Education and Research (NISER), HBNI,, P.O.- Bhimpur, Padanpur, Via-Jatni, 752050, India}

\author{Shaktiranjan Mohanty}
\affiliation{Laboratory for Nanomagnetism and Magnetic Materials (LNMM), School of Physical Sciences, National Institute of Science Education and Research (NISER), HBNI,, P.O.- Bhimpur, Padanpur, Via-Jatni, 752050, India}

\author{Subhankar Bedanta}
\email{sbedanta@niser.ac.in}
\affiliation{Laboratory for Nanomagnetism and Magnetic Materials (LNMM), School of Physical Sciences, National Institute of Science Education and Research (NISER), HBNI,, P.O.- Bhimpur, Padanpur, Via-Jatni, 752050, India}
\affiliation{Center for Interdisciplinary Sciences (CIS), NISER, HBNI, P.O.-Bhimpur Padanpur, Jatni, Khurda, 752050, India}
\date{\today}

\begin{abstract}
\section*{ABSTRACT}

Transition metal dichalcogenides (TMD) possess novel properties which makes them potential candidates for various spintronic applications. Heterostructures of TMD with magnetic thin film have been extensively considered for spin-orbital torque, enhancement of perpendicular magnetic anisotropy etc. However, the effect of TMD on magnetic anisotropy in heterostructures of in-plane magnetization has not been studied so far. Further the effect of the TMD on the domain structure and magnetization reversal of the ferromagnetic system is another important aspect to be understood. In this context we study the effect of MoS$_2$, a well-studied TMD material, on magnetic properties of CoFeB in MoS$_2$/CoFeB heterostructures. The reference CoFeB film possess a weak in-plane anisotropy. However, when the CoFeB is deposited on MoS$_2$ the in-plane anisotropy is enhanced as observed from magneto optic Kerr effect (MOKE) microscopy as well as ferromagnetic resonance (FMR). Magnetic domain structure and magnetization reversal have also been significantly modified for the MoS$_2$/CoFeB bilayer as compared to the reference CoFeB layer. Frequency and angle dependent FMR measurement show that the magnetic anisotropy of CoFeB increases with increase in thickness of MoS$_2$ in the MoS$_2$/CoFeB heterostructures.

\end{abstract}


\maketitle
Two-dimensional (2D) materials are composed of atomically thin layers that are bonded covalently or ionically along in-plane and Van der Waals along out-of-plane to the layer. Graphene was discovered first among the 2D materials and it is well known for its various exotic properties\cite{dash2014graphene,ando2007exotic,castro2009electronic}. Transition metal dichalcogenides (TMDs) are another class of 2D materials of the form MX$_2$ with M and X as transition metals and chalcogen group elements (S, Se, Ta), respectively\cite{wang2012electronics}. Here each layers are composed of covalently sandwiched transition metal in between chalcogen group elements. These layers are stacked repeatedly through Van der Waals bonds between them.These materials can exhibit insulating, conducting, semiconducting to superconducting properties\cite{chhowalla2013chemistry}. 
Unlike graphene, TMDs have electronic band gap and high SOC\cite{wang2012electronics,chhowalla2013chemistry,guguchia2018magnetism}. However, most of them are inherently nonmagnetic at room temperature which restricts their usage for spintronics and memory devices based applications \cite{ng2020high,liu2020spintronics}. Recently, number of techniques such as magnetic impurity doping \cite{lin2016magnetism}, defect engineering \cite{guguchia2018magnetism,ge2021magnetic,lin2016defect}, atomic intercalation \cite{ahamd2020intercalation} and heterostructures \cite{zhong2017van,zhang2019ferromagnet} have been reported to induce and subsequently manipulate magnetism in TMDs. Among these, heterostructures which comprise of TMD and ferromagnetic (FM) materials have drawn attention because of various spin based phenomena such as anomalous Hall effect \cite{ng2020high}, Rashba spin-orbit coupling \cite{liu2020spintronics,wang2015physical}, quantum spin Hall effect \cite{liu2020spintronics},spin-valley locking \cite{marfoua2020electric}, perpendicular magnetic anisotropy \cite{zhang2019ferromagnet}, spin injection \cite{garandel2017electronic} and spin orbit torque (SOT) \cite{hidding2020spin}. All these spin based phenomena in TMD/FM heterostructures are due to the combined effect of high SOC in TMD and spin polarization behaviour of FM. Apart from these, enhancement of perpendicular magnetic anisotropy has also been reported in MoS$_2$ based TMD/FM heterostructures and it is explained to be due to interfacial hybridization\cite{xie2019giant}. However,  such interfacial effect in a in-plaine magnetic anisotropic system and its magnetic domain structures has not been investigated. Therefore, the focus of this paper is to study the effect of MoS{$_2$} on magnetization reversal, domain structures, anisotropy of CoFeB in MoS{$_2$}/CoFeB heterostructures. In this paper, we show that the MoS$_2$ induces an uniaxial in-plane magnetic anisotropy and significantly affect the domain structure of CoFeB (CFB).

A high vacuum multi-deposition chamber (Make: Mantis Deposition Ltd., UK) with base pressure $\sim$ 10$^{-8}$ mbar was used to prepare the thin films. 
MoS$_2$ and CoFeB (Co$_{40}$Fe$_{40}$B$_{20}$) films were deposited from their corresponding commercially available stoichiometric targets. CoFeB and MoS$_2$ layers were deposited by DC and RF magnetron sputtering, respectively, in Ar environment. A capping layer of MgO ($\sim$ 2 nm) to prevent the CFB layer from ambience was deposited on top of all the samples using electron beam evaporation. A reference sample Si/SiO{$_2$}/CFB(6nm)/MgO has been prepared which is designated as sample M0. Two heterostructure samples Si/SiO{$_2$}/MoS{$_2$}($t$ nm)/CFB(6nm)/MgO with $t$ $=$ 6 and 28 nm have been prepared which are designated as samples M1 and M2, respectively. A schematic of sample structure is shown in figure 1 (a). The oxidized Si substrate (Si/SiO{$_2$}) was fabricated by thermal oxidation of commercial Si substrate at 1000$^{\circ}$C in oxygen atmosphere for 10 hours. The deposition pressure for MoS$_2$ and CFB were 5.5 \textit{$\times$}10$^{-3}$ and 8\textit{$\times$} 10$^{-4}$ mbar, respectively. The deposition rate was fixed to 0.1 {\AA}/s for both MoS$_2$ and CFB. During deposition, all three samples were rotated with 20 rotation per minute (RPM) to suppress growth dependent anisotropy as well as to establish a homogeneous deposition.
One of the heterostructure samples M2 was characterized using transmission electron microscope (TEM) (Make: JEOL JEM F200). Presence of MoS$_2$ in the samples M1 and M2 were characterized using laser Raman spectroscopy (Make: HORIBA LabRAM HR Evolution) with laser wavelength of 532 nm. 
 
 Magnetization reversal of all three samples were studied by measuring magnetic hysteresis loops along with simultaneous domain imaging using magneto optic Kerr effect (MOKE) based microscope (Make: Evico Magnetics Ltd. Germany) in longitudinal geometry. Angle dependence of magnetization reversal was studied by applying magnetic field at different angles with respect to the easy axis of the sample at an angular interval of \textit{$\phi$} $=$ 10$^{\circ}$.  In order to measure damping constant and anisotropy field, frequency dependent ferromagnetic resonance (FMR) (Make: NanOsc Instrument Phase FMR) was performed by sweeping the magnetic field from 200 to 2000 Oe at different fixed RF-field frequencies between 5 and 17 GHz with an interval of 0.5 GHz. Further in order to quantify the in-plane anisotropy angle dependent FMR was performed by rotating the sample at a fixed RF frequency of 12 GHz. 
 
  \begin{figure}[ht]
	\centering
	\includegraphics[width=0.48\textwidth]{"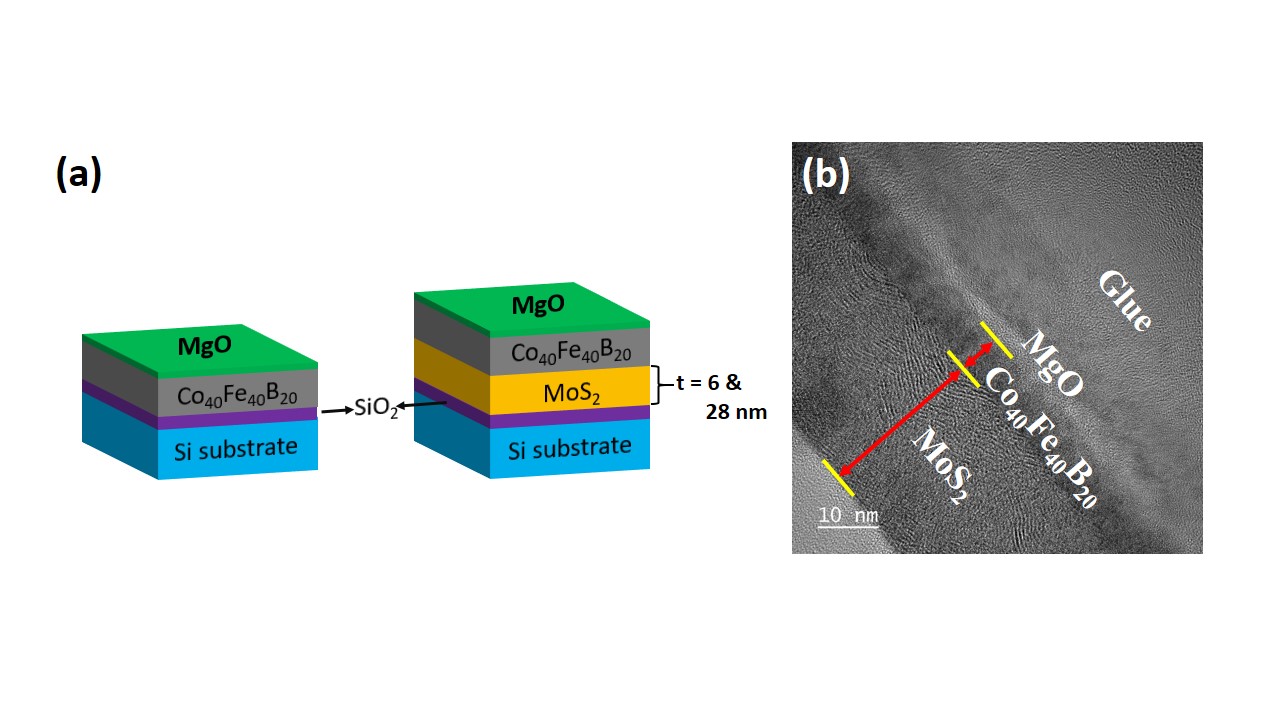"}
	\caption{(a) Schematic sample structure of CFB reference sample and MoS$_{2}$/CFB heterostructure on Si/SiO$_{2}$ substrate (b) Cross-sectional TEM image of sample M2 shows clear contrast between SiO$_{2}$, MoS$_{2}$, CFB and MgO}
	\label{fig:figure-1}
\end{figure}
 
 \begin{figure}[ht]
	\centering
	\includegraphics[width=0.48\textwidth]{"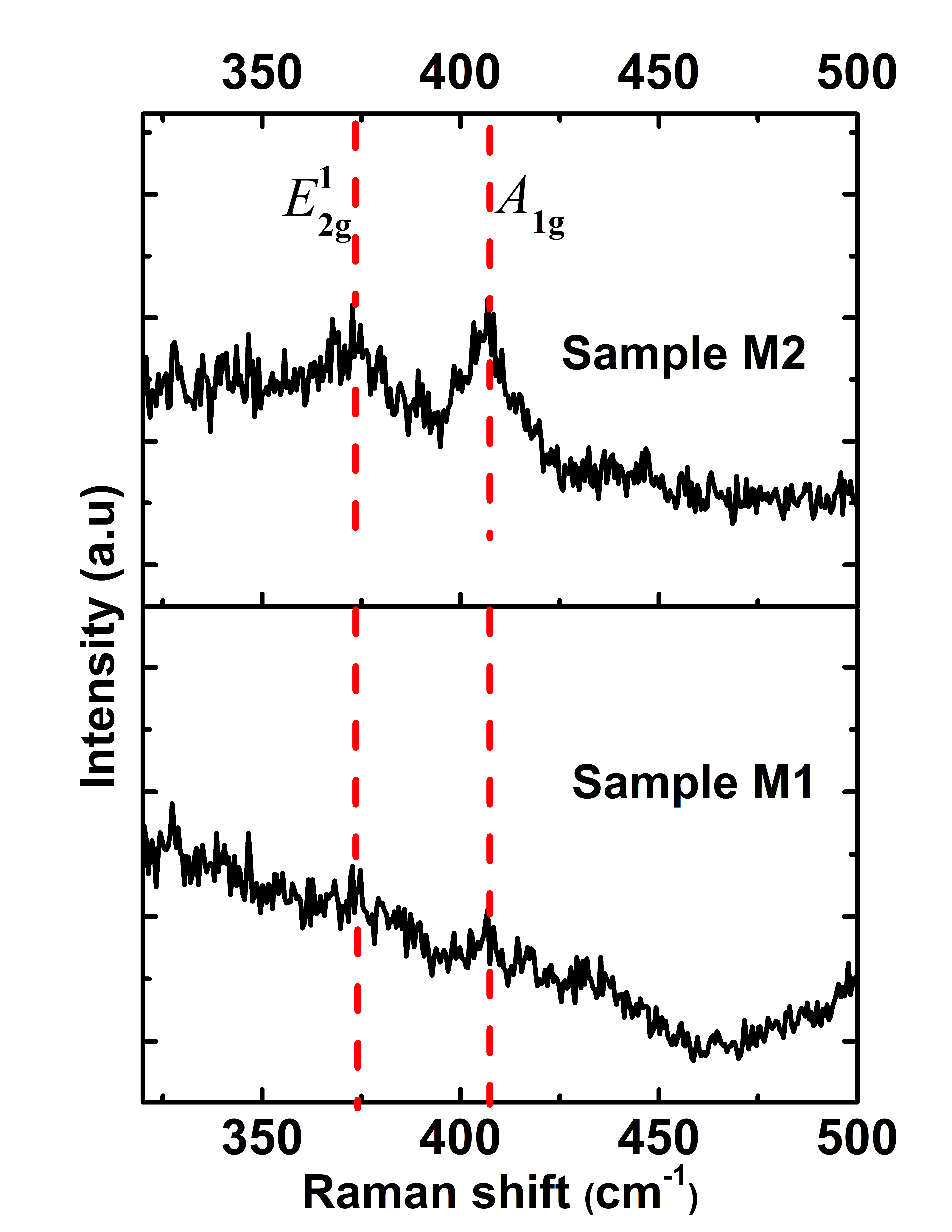"}
	\caption{Raman spectra of sample M1 and M2 }
	\label{fig:figure-2}
\end{figure}

  A cross-sectional TEM image of sample M2 shown in figure 1(b) exhibits clear contrast between MoS$_{2}$, CFB and MgO layers on top of SiO$_{2}$. It shows sharp interfaces between the layers and indicates absence of interdiffusion. Raman spectra of samples M1 and M2 are shown in figure 2. Sample M2 exhibits Raman peaks corresponding to in-plane ($\text{E}^1_{2g}$) and out-of-plane ($\text{A}_{1g}$) characteristic vibrational modes for MoS{$_2$} at $\mathtt{\sim}$ 382 cm$^{-1}$ and 408 cm$^{-1}$, respectively. The detailed Raman spectra analysis for similar MoS$_2$ samples can be found in the supplementary information.

\begin{figure}[ht]
	\centering
	\includegraphics[width=0.48\textwidth]{"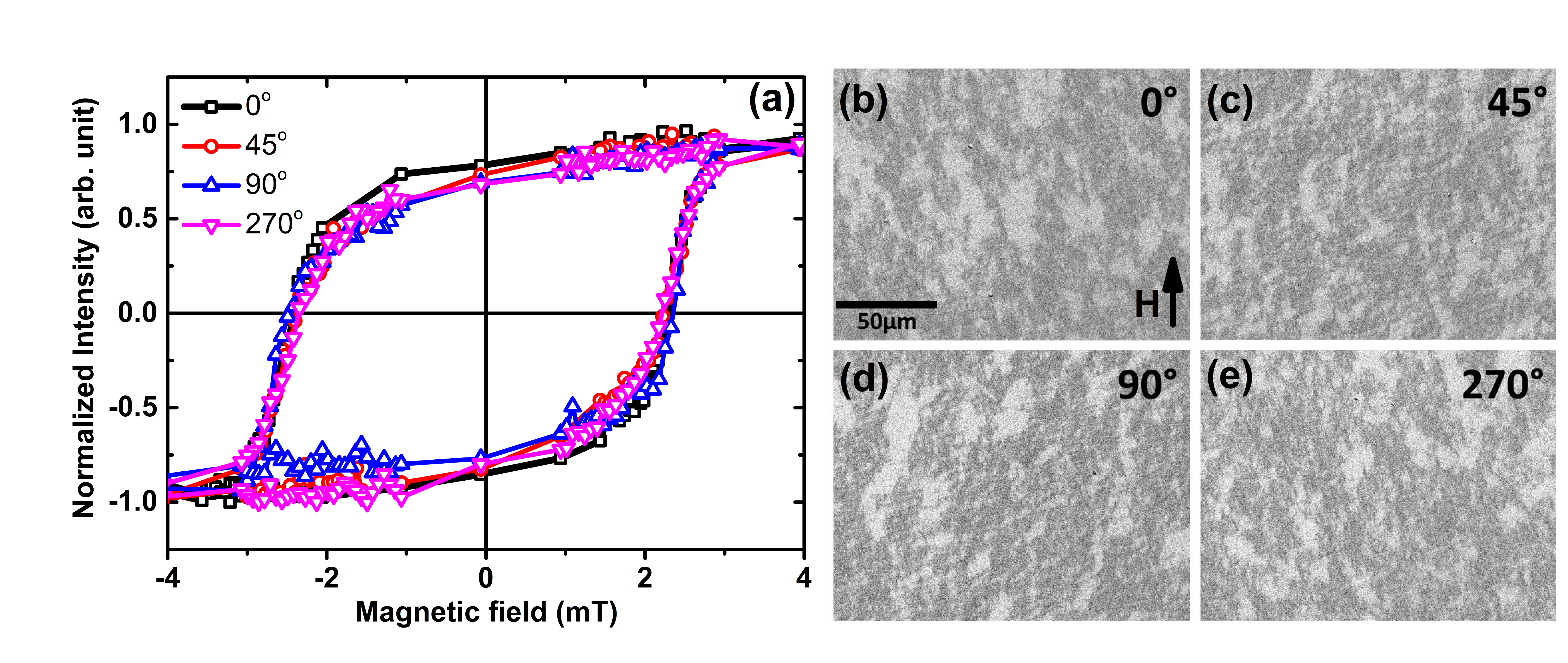"}
	\caption{(a) Hysteresis loops measured at angles \textit{$\phi$} = 0$^{\circ}$, 45$^{\circ}$, 90$^{\circ}$ and 270$^{\circ}$ for sample M0 and (b, c, d and e) corresponding domain images captured near to coercive field using MOKE based microscope in longitudinal geometry at room temperature. Scale bar and magnetic field direction shown in (b) are valid for all remaining domain images}
	\label{fig:figure-3}
\end{figure}

  Angle dependence of magnetic hysteresis loops and domain images captured at coercive field for sample M0 are shown in figure 3. Domain images show similar patch like domains at all angles of measurement. Hysteresis loops as well as the domain structures are independent of angles at which it was measured. This hints the isotropic behaviour of CFB layer.The isotropic or anisotropic nature of a magnetic film can be estimated using orientation ratio (OR), a ratio of remanent magnetization measured along easy axis to hard axis (\textit{$M_r^{EA}$}⁄\textit{$M_r^{HA}$}) \cite{arnoldussen1984obliquely}. The value of OR for ideal isotropic and anisotropic magnetic system is one and infinite, respectively \cite{arnoldussen1984obliquely,johnson1995plane}. The calculated value of OR for the sample M0 is 0.99. These results confirm that the CFB layer is magnetically isotropic in the film plane, a behavior inherent to amorphous ferromagnets where the magnetocrystalline anisotropy is insignificant due to lacking in long-range structural order \cite{xie2020magnetocrystalline,nayar1990structural}. 
\begin{figure}[ht]
	\centering
	\includegraphics[width=0.48\textwidth]{"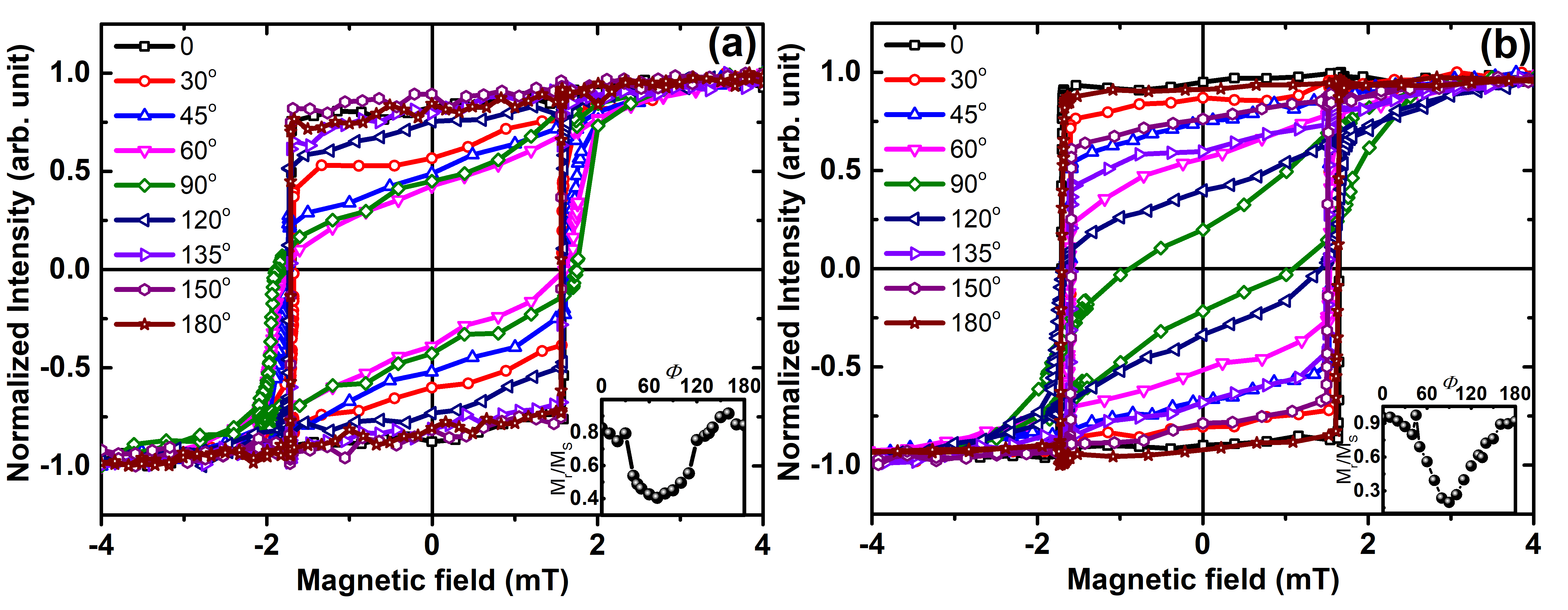"}
	\caption{Hysteresis loops measured at angles \textit{$\phi$} = 0$^{\circ}$ to 180$^{\circ}$ for sample M1 (a) and M2 (b) using MOKE based microscope in longitudinal geometry at room temperature. Inset shows the angle dependence of \textit{$M_r/M_s$}.}
	\label{fig:figure-4}
\end{figure}
 On the other hand, the hysteresis loops and domain images of samples M1 and M2 (figure 4) exhibit different behavior as compared to the reference sample M0. The coercivity changes with \textit{$\phi$} which indicates presence of anisotropy. The angle dependence of \textit{$M_r$}/\textit{$M_s$} ratio shown for both samples in the inset of the figure 4 (a) and (b) indicates the existence of uniaxial anisotropy with maximum and minimum value at \textit{$\phi$} $=$ 0$^{\circ}$ and 90$^{\circ}$, respectively. In addition, a sharp variation in angle dependence of \textit{$M_r$}/\textit{$M_s$} ratio exhibited by sample M2 implies that the anisotropy dispersion in sample M2 is lesser than that in sample M1.  For an ideal uniaxial anisotropic system with minimal anisotropy dispersion, the squareness ratio of hysteresis loop measured along easy axis \textit{$[{M_r/M_s}]^{ea}$} and the coercivity of hysteresis loop measured along hard axis \textit{${H}_c^{ha}$} are close to 1 and 0, respectively\cite{idigoras2011collapse,chowdhury2018360, mallick2018tuning}. Deviation of \textit{$[{M_r/M_s}]^{ea}$} and \textit{${H}_c^{ha}$} from these values is the measure of extent to which the anisotropy is dispersed in the system\cite{beach2005co}. The values of \textit{$[{M_r/M_s}]^{ea}$}, OR and \textit{${H}_c^{ha}$} for samples M1 and M2 are shown in table 1. \textit{$[{M_r/M_s}]^{ea}$} and OR increases and \textit{${H}_c^{ha}$} decreases with increase in thickness of MoS$_2$. These results confirm the decrease in anisotropy dispersion in CFB layer with increase in thickness of MoS$_2$.

 \begin{table}[htp]
\caption{ \textit{$[{M_r/M_s}]^{ea}$} (Dimensionless),OR (Dimensionless) and $H_{c}^{ha}$ (mT) value of sample M1 ans M2} \label{tab:title} 
\begin{center}
	\begin{tabular}{c c c c} 
		\hline
		Sample & \textit{$[{M_r/M_s}]^{ea}$} & OR & $H_{c}^{ha}$ \\
		\hline
		M1 & 0.83 & 1.8 & 1.81 \\
		\hline
		M2 & 0.95 & 4.8 & 0.98 \\ [1ex]
		\hline		
	\end{tabular}
\end{center}
\end{table}

\begin{figure}[htp]
	\centering
	\includegraphics[width=0.48\textwidth]{"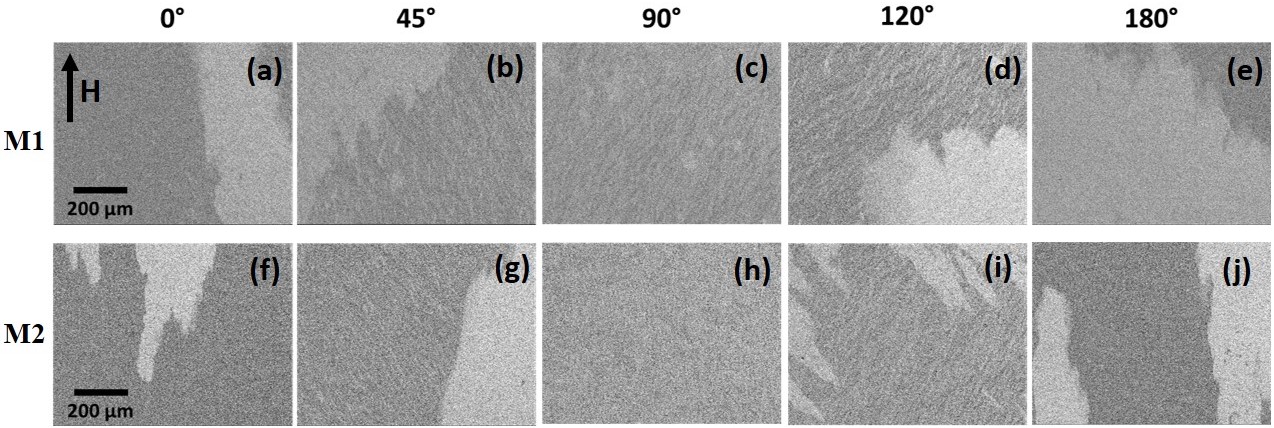"}
	\caption{Domain images captured for samples M1 and M2 at angles \textit{$\phi$} = 0$^{\circ}$, 45$^{\circ}$, 90$^{\circ}$, 120$^{\circ}$ and 180$^{\circ}$ near to coercivity. Scale bar and magnetic field direction shown in (a) are valid for all remaining images. }
	\label{fig:figure-5}
\end{figure}
 The domain images for the samples M1 and M2 captured near to ${H}_c$ measured at different angles are shown in figure 5. The domain images along 0$^{\circ}$ (figure 5 (a) and (f)) and 180$^{\circ}$ (figure 5 (e) and (j)) show large domains which indicate that the magnetization reversal occurs via domain wall nucleation and motion \cite{chowdhury2014controlling}. Shape of these large domains are random (figure 5 (a) and (e)) for the sample M1 whereas it is well defined stripes (figure 5 (f) and (j)) in the case of sample M2. At 90$^{\circ}$, fine ripple domains for the sample M1 (figure 5 (c)) and no domains for the sample M2 (figure 5 (h)) indicate that the magnetization reversal occurs by coherent rotation \cite{mallik2018study}. In between easy and hard axis, that is for the angles 0$^{\circ}$ $<$ \textit{$\phi$} $<$ 90$^{\circ}$ and 90$^{\circ}$ $<$ \textit{$\phi$}  $<$ 180$^{\circ}$, a narrow ripple domains (figure 5 (b), (d), (g) and (i)) were observed along with big stripe domains for the samples M1 and M2, respectively. Further to understand the evolution of these two types of domain structures, domain reversal videos SV1 to SV4 captured during magnetization reversal at \textit{$\phi$} $=$ 45$^{\circ}$ and 120$^{\circ}$ for both the samples M1 and M2 are provided as supplementary information. All the four domain reversal videos show that the magnetization reversal occurs via two distinct steps: (i) appearance of fine ripple domains followed by (ii) annihilation of ripple domains by nucleation and wall motion of big stripe domains. 
 For a conventional magnetic thin film with nominal anisotropy dispersion, the field required for domain wall motion $H_W$ is lesser than that for magnetization rotation $H_R$. In other words, the ratio $H_W$/$H_R$ $<$ 1. Therefore, in this case the magnetization reversal occurs via domain wall motion rather than coherent rotation. On the other hand, the ratio $H_W$/$H_R$ is inverted,i.e, $H_W$/$H_R$ $<$ 1 along all field directions for a film fabricated with large anisotropy dispersion. These films are specially called as “inverted films”. \cite{smith1960anisotropy,methfessel1961partial,cohen1962anomalous,cohen1963influence}. It has also been reported that the films with moderate dispersion in anisotropy exhibit the inversion behaviour in a sector of field directions between easy and hard axis \cite{methfessel1961partial}. From this it can be clearly understood that the samples M1 and M2 are the films having moderate dispersion in anisotropy. Prominence of ripple domain in the case of sample M1 indicates a larger anisotropy dispersion in sample M1 than that in sample M2.
 
 Further to quantify the anisotropy and the symmetry, frequency and angle dependent FMR spectra were performed on samples M1 and M2. For each excitation frequency, the resonance field \textit{$H_{res}$} and the line width \textit{$\Delta H$} have been extracted by fitting the FMR derivative signal to the Lorentzian function;
 
 \begin{multline}
 S_{21}=A_1\frac{4\Delta H(H-H_{res})}{[4(H-H_{res})^2 + {(\Delta H)}^2]})+\\ A_2\frac{{(\Delta H)}^2-4{(H-H_{res})}^2}{[4(H-H_{res})^2 + {(\Delta H})^2]})+Slope H+Offset
\end{multline}

 Where slope \textit{$H$} is drift value in amplitude of the signal, \textit{$A_1$} and \textit{$A_2$} are coefficient of anti-symmetric and symmetric components, respectively. 
 
 \begin{figure}[ht]
	\centering
	\includegraphics[width=0.48\textwidth]{"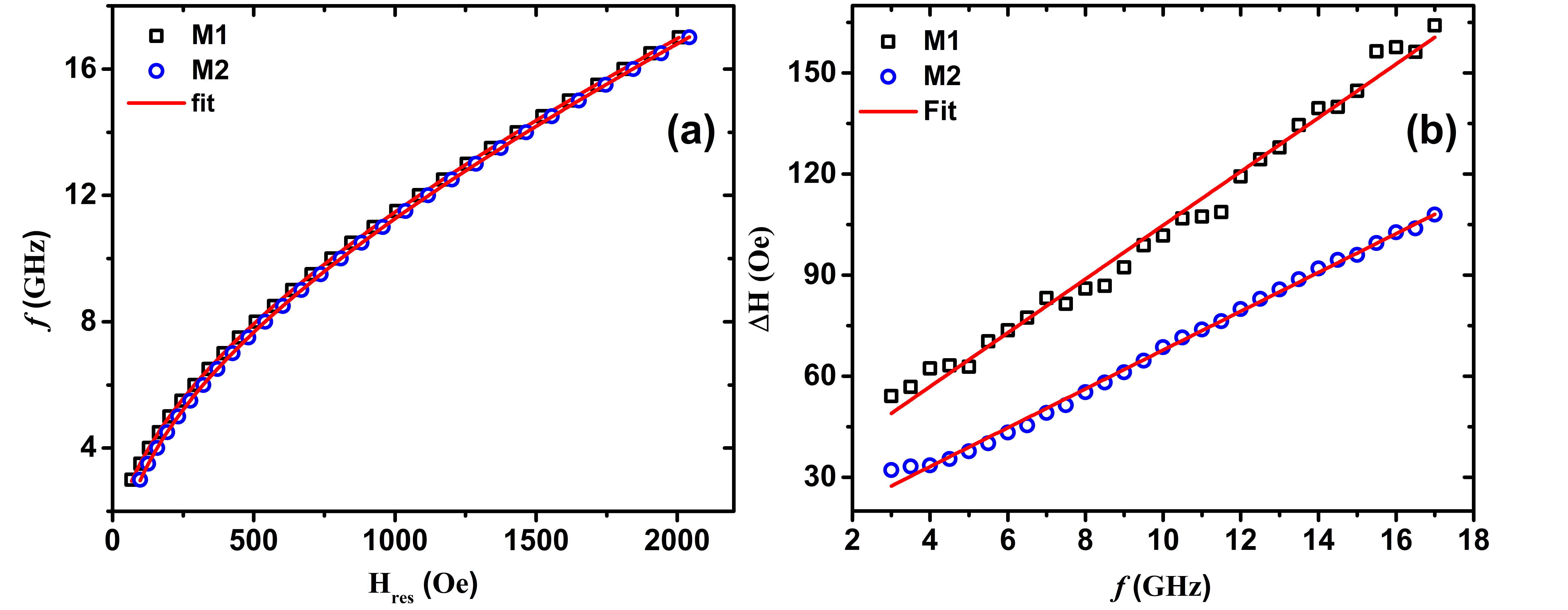"}
	\caption{(a) \textit{$f$} vs \textit{$H_{res}$} and (b) \textit{$\Delta H$} vs \textit{$f$} plots for samples M1 and M2 extracted from the frequency dependent FMR spectra. The solid lines in (a) and (b) are the best fits to equations (2) and (3) respectively.}
	\label{fig:figure-6}
\end{figure}
 
 Dependence of \textit{$H_{res}$} and \textit{$\Delta H$} on frequency \textit{f} are plotted in figures 6 (a) and (b), respectively. Resonance condition for \textit{f} as a function of $\text{H}_{res}$ is given by Kittel equation as \cite{kittel1948theory}.
 \begin{equation}
f=\frac{\gamma}{2\pi}\sqrt{(H_{res}+H_{K})(H_{res}+4\pi M_{eff}+H_{K})}
\label{Kittel equation}
\end{equation}
 
 Where \textit{${H}_K$} is in-plane anisotropy field, $4\pi M_{eff}$ is effective magnetization and \textit{$\gamma=g\mu_B/\hbar$} is gyromagnetic ratio with \textit{g} as Lande spectroscopic splitting factor, \textit{$\mu_B$} as Bohr magneton, \textit{$\hbar$} as reduced Planck’s constant. Value of ${H}_K$, $4\pi M_{eff}$ and \textit{g} were extracted from the fit of the \textit{f} vs ${H}_{res}$ plot (figure 6(a)) to the Kittel resonance condition (equation 2). The FMR line-width \textit{$\Delta  H$} can be written in terms of broadening due to magnetic inhomogeneity \textit{$\Delta  H_0$} and intrinsic Gilbert-damping constant \textit{$\alpha$} of the sample is as follows \cite{nembach2011perpendicular};
\begin{equation}
\Delta H=\Delta H_{0}+\dfrac{4\pi\alpha f}{\gamma} .
\label{non linear equation fit}
\end{equation} 
 
 Value of \textit{$\Delta  H_0$} and \textit{$\alpha$} were extracted from the fit of the \textit{f} vs \textit{$\Delta  H$} plot (figure 6b) to the equation 3. The values of ${H}_K$, $4\pi M_{eff}$, \textit{g}, \textit{$\alpha$} and \textit{$\Delta  H_0$} extracted using equations 2 and 3 for samples M1 and M2 are shown in table 2. 
 
  \begin{table}[H]
\caption{Parameters extracted from the fitting of FMR experimental data (figure 6) of both the samples M1 and M2 using equations (2) and (3).} \label{tab:title} 
\begin{center}
	\begin{tabular}{c c c c c c} 
		\hline
		Sample & \textit{H$_K$}(Oe) & $4\pi M_{eff}$(Oe) & \textit{g} & \textit{$\alpha$} & \textit{$\Delta H_0$}(Oe)  \\ [1ex]
		\hline
		M1 & 5.449 & 9271.40 & 2.54 & 0.014 & 25.02 \\ [1ex]
		\hline
		M2 & -23.31 & 9027.45 & 2.56 & 0.010 & 10.13 \\ [1ex]
		\hline		
	\end{tabular}
\end{center}
\end{table}
 
The value of \textit{g} obtained for both the samples are very large and this may be due to the major contribution from the orbital moments. The detailed discussion about the high value of \textit{g} is made in the supplementary information. 
It is found that with increase in thickness of MoS$_2$, the value of \textit{$H_K$} increases whereas the value of \textit{$\Delta  H_0$} decreases. This indicates that the increase in anisotropy strength and decrease in anisotropy
dispersion occur with increase in thickness of MoS$_2$. These results are in consistent with results obtained from Kerr microscopy. Further the anisotropy and symmetry of samples M0, M1 and M2 were quantified from angle dependent FMR measurement. For sample M0, a weak \textit{$\phi$} dependence of \textit{$H_{res}$} in \textit{$\phi$} vs \textit{$H_{res}$} plot (not shown here) implies the existence of large anisotropy dispersion.

\begin{figure}[ht]
	\centering
	\includegraphics[width=0.48\textwidth]{"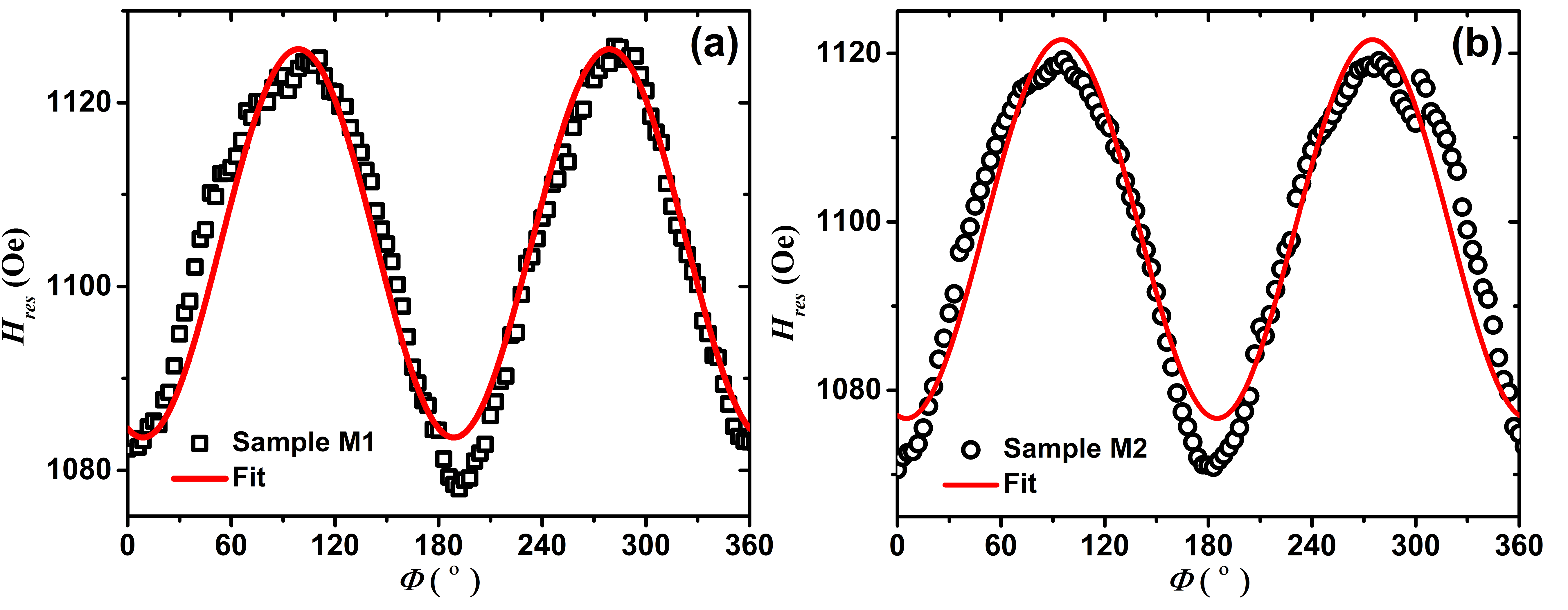"}
	\caption{Angle dependent \textit{$H_{res}$} plot measured using FMR by keeping the frequency fixed at 12 GHz for for samples (a) M1 and (b) M2. The solid lines are the best fits to equation (4).}
	\label{fig:figure-7}
\end{figure}

This is in agreement with the Kerr microscopy results of sample M0 which confirms a magnetically isotropic (in the film plane) behaviour. \textit{$\phi$} dependence of \textit{$H_{res}$} for samples M1 and M2 in figure 7 clearly show an uniaxial magnetic anisotropy with \textit{$H_{res}$} maximum at $\mathtt{\sim}$    90$^{\circ}$ and 270$^{\circ}$ and minimum at ~ 0$^{\circ}$, 180$^{\circ}$ and 360$^{\circ}$. The dispersion relation for the resonance condition in the in-plane geometry is given by \cite{pan2017role} 
 \begin{multline}
 f=\frac{\gamma}{2\pi}([H+\frac{2K}{M_s}\cos{2\phi}]\times[H+4\pi M_s+\frac{2K}{M_s}\cos{2\phi}])
\end{multline}
 
 Where, \textit{$K$} and \textit{$M_{s}$} are in-plane anisotropy and saturation magnetization, respectively. Value of \textit{$K$} and \textit{$M_{s}$} were extracted by fitting of the \textit{$\phi$} dependence of \textit{$H_{res}$} to the equation 4 and shown in table 3. The value of in-plane anisotropy increases with increase in thickness of MoS$_2$. A plausible reason for the induced uniaxial anisotropy in CFB layer by MoS$_2$ could be the orbital hybridisation accompanied by the charge transfer at their interface. At the MoS$_2$/CFB interface, a d-d orbital hybridization is possible due to more than half filled d-band of both Co (3d$^7$) and Fe (3d$^6$) and less than half filled d-band of Mo$^{4+}$ (4d$^1$) in MoS$_2$ \cite{xie2019giant,zhang2019ferromagnet}.

  \begin{table}[H]
\caption{Parameters extracted from the fitting of angle dependent FMR experimental data (figure 7) of both samples using equation (4)} \label{tab:title} 
\begin{center}
	\begin{tabular}{c c c} 
		\hline
		Sample & \textit{$K$}(erg/cm$^3$) & \textit{$M_{s}$}(emu/cm$^3$) \\
		\hline
		M1 & -8081.6 & 727.6 \\
		\hline
		M2 & 8406.7 & 711.0 \\ [1ex]
		\hline		
	\end{tabular}
\end{center}
\end{table}

 In this paper, effect of MoS$_2$ on magnetization reversal, domain structure, magnetic anisotropy of CoFeB was studied using MOKE based microscopy and FMR spectroscopy. Magnetization reversal and corresponding domain structures show that when CoFeB is being deposited on top of the MoS$_2$, due to hybridization between MoS$_2$ and Co or Fe, the anisotropy is getting enhanced. This enhancement in anisotropy also leads to change in magnetic domain structure and reversal process. Further the domain structure reveals the inverted film behaviour, an indication of existence of dispersion in anisotropy. Frequency and angle dependent FMR measurement show that for increase in thickness of MoS$_2$ increase the magnetic anisotropy and decrease the magnetic inhomogeneity.

We acknowledge the financial support by the Department of Atomic Energy (DAE), the Department of Science and Technology
(DST-SERB) of Government of India, the DST-Nanomission [project sanction No.
SR/NM/NS-1018/2016(G)]. We thank Dr. B. B. Singh for the valuable discussions related to the oxidation of Si substrates.  

\section*{DATA AVAILABILITY}
The data that support the findings of this study are available from the corresponding author upon request.

\section*{REFERENCES}
\bibliographystyle{apsrev4-2}
\bibliography{manuscript}

\end{document}